\documentstyle[aps,preprint,floats,epsf,epsfig]{revtex}
\begin{document}
\def\be{\begin{eqnarray}}
\def\en{\end{eqnarray}}
\def\non{\nonumber}
\def\la{\langle}
\def\ra{\rangle}
\def\nc{{N_c^{\rm eff}}}
\def\vp{\varepsilon}
\def\drho{\bar\rho}
\def\deta{\bar\eta}
\def\vma{{_{V-A}}}
\def\vpa{{_{V+A}}}
\def\J{{J/\psi}}
\def\ov{\overline}
\def\Lqcd{{\Lambda_{\rm QCD}}}
\def\pr{{\sl Phys. Rev.}~}
\def\prl{{\sl Phys. Rev. Lett.}~}
\def\pl{{\sl Phys. Lett.}~}
\def\np{{\sl Nucl. Phys.}~}
\def\zp{{\sl Z. Phys.}~}
\def\lsim{ {\ \lower-1.2pt\vbox{\hbox{\rlap{$<$}\lower5pt\vbox{\hbox{$\sim$}
}}}\ } }
\def\gsim{ {\ \lower-1.2pt\vbox{\hbox{\rlap{$>$}\lower5pt\vbox{\hbox{$\sim$}
}}}\ } }

\font\el=cmbx10 scaled \magstep2{\obeylines \hfill May, 2001}

\vskip 1.5 cm

\centerline{\large\bf Charmless $B\to VV$ Decays in QCD
Factorization:} \centerline{\large\bf Implications of Recent $B\to
\phi K^*$ Measurement}
\bigskip
\centerline{\bf Hai-Yang Cheng$^{1,2}$ and Kwei-Chou Yang$^{3}$}
\medskip
\centerline{$^1$ Institute of Physics, Academia Sinica}
\centerline{Taipei, Taiwan 115, Republic of China}
\medskip
\centerline{$^2$ Physics Department, Brookhaven National
Laboratory} \centerline{Upton, New York 11973}
\medskip
\centerline{$^3$ Department of Physics, Chung Yuan Christian
University} \centerline{Chung-Li, Taiwan 320, Republic of China}
\bigskip
\bigskip
\centerline{\bf Abstract}
\bigskip
{\small In the heavy quark limit, both vector mesons in the
charmless $B\to VV$ decay should have zero helicity and the
corresponding amplitude is proportional to the form factor
difference $(A_1-A_2)$. The first observed charmless $B\to VV$
mode, $B\to\phi K^*$, indicates that the form factors $A_1(q^2)$
and $A_2(q^2)$ cannot be very similar at low $q^2$ as shown in
some form-factor models. The approach of QCD-improved
factorization implies that the nonfactorizable correction to each
partial-wave or helicity amplitude is not the same; the effective
parameters $a_i$ vary for different helicity amplitudes. The
leading-twist nonfactorizable corrections to the transversely
polarized amplitudes vanish in the chiral limit and hence it is
necessary to take into account twist-3 distribution amplitudes of
the vector meson in order to have renormalization scale and scheme
independent predictions. Branching ratios of $B\to VV$ decays are
calculated in two different models for form factors, and the
predicted decay rates are different by a factor of $1.5\sim 2$.
Owing to the absence of $(S-P)(S+P)$ penguin contributions to the
$W$-emission amplitudes, tree-dominated decays tend to have larger
branching ratios than the penguin-dominated ones.

} \pagebreak

{\bf 1.}~~It is known that the decay amplitude of a $B$ meson into
two vector mesons  is governed by three unknown form factors
$A_1(q^2),~A_2(q^2)$ and $V(q^2)$ in the factorization approach.
It has been pointed out in \cite{CCTY} that the charmless $B\to
VV$ rates are very sensitive to the form-factor ratio $A_2/A_1$.
This form-factor ratio is almost equal to unity in the
Bauer-Stech-Wirbel (BSW) model \cite{BSW}, but it is less than
unity in the light-cone sum rule (LCSR) analysis for form factors
\cite{Ball}. In general, the branching ratios of $B\to VV$
predicted by the LCSR are always larger than that by the BSW model
by a factor of $1.6\sim 2$ \cite{CCTY}. This is understandable
because in the heavy quark limit, both vector mesons in the
charmless $B\to VV$ decay should have zero helicity and the
corresponding amplitude is proportional to the form factor
difference $(A_1-A_2)$. These two form factors are identical at
$q^2=0$ in the BSW model. We shall see that the first observed
charmless $B\to VV$ mode, $B\to\phi K^*$, recently measured by
CLEO \cite{Briere}, BELLE \cite{BELLE} and BABAR \cite{BaBar},
clearly favors the LCSR over the BSW model for $B-V$ transition
form factors.

In the present paper we will embark on a study of $B\to VV$ decays
in the approach of QCD-improved factorization which enables us to
compute nonfactorizable corrections in the heavy quark limit. In
the so-called generalized factorization approach, it is assumed
that nonfactorizable effects contribute to all partial-wave or
helicity amplitudes in the same weight. We shall see that it is
not the case in QCD factorization. Moreover, we will show that the
leading-twist nonfactorizable corrections to the transversely
polarized amplitudes vanish in the chiral limit. Hence it is
necessary to go beyond the leading-twist approximation for
transversely polarized states.

\vskip 0.7cm
{\bf 2.}~~In general the $B\to VV$ amplitude consists
of three independent Lorentz scalars:
 \be A[B(p)\to
V_1(\vp_1,p_1)V_2(\vp_2,p_2)] \propto \vp_1^{*\mu}\vp_2^{*\nu}(a
g_{\mu\nu}+bp_\mu p_\nu+ic\epsilon_{\mu\nu\alpha\beta}p_1^\alpha
p_2^\beta), \label{amp}
 \en
where the coefficient $c$ corresponds to the $p$-wave amplitude,
and $a$, $b$ to the mixture of $s$- and $d$-wave amplitudes. Three
helicity amplitudes can be constructed as
 \be
H_{00} &=& {1\over
2m_1m_2}\left[ (m_B^2-m_1^2-m_2^2)a+2m_B^2p_c^2b\right],  \non \\
H_{\pm\pm} &=& a\mp m_Bp_cc,
 \en
where $p_c$ is the c.m. momentum of the vector meson in the $B$
rest frame and  $m_1$ ($m_2$) is the mass of the vector meson
$V_1$ ($V_2$). For $H_{--}$ to occur quark spin in the emitted
vector meson $V_2$ has to be flipped. Therefore, the amplitude
$H_{--}$ is suppressed by a factor of $m_2/m_B$ \cite{Korner}. The
$H_{++}$ amplitude is subject to a further chirality suppression
of order $m_1/m_B$. In general, it is thus expected that
$|H_{00}|^2>|H_{--}|^2>|H_{++}|^2$. The total decay rate is given
by
\be
\Gamma(B\to V_1V_2)=\,{p_c\over 8\pi
m_B^2}(|H_{00}|^2+|H_{++}|^2+|H_{--}|^2).
\en

In terms of the decay constant and form factors defined by
\cite{BSW}:
\be
\la V(p,\vp)|V_\mu|0\ra &=& f_Vm_V\vp^*_\mu,  \non \\ \la
V(p',\vp)|V_\mu|P(p)\ra &=& {2\over
m_P+m_V}\,\epsilon_{\mu\nu\alpha \beta}\vp^{*\nu}p^\alpha p'^\beta
V(q^2),   \non \\ \la V(p',\vp)|A_\mu|P(p)\ra &=& i\Big\{
(m_P+m_V)\vp^*_\mu A_1(q^2)-{\vp^*\cdot p\over
m_P+m_V}\,(p+p')_\mu A_2(q^2)    \non
\\ && -2m_V\,{\vp^*\cdot p\over
q^2}\,q_\mu\big[A_3(q^2)-A_0(q^2)\big]\Big\},
\en
where $q=p-p'$, $A_3(0)=A_0(0)$, and
\be
A_3(q^2)=\,{m_P+m_V\over 2m_V}\,A_1(q^2)-{m_P-m_V\over
2m_V}\,A_2(q^2),
\en
one has the factorizable $B\to V_1V_2$ amplitude:
\be
X^{( BV_1,V_2)} &\equiv & \la V_2 | (\bar{q}_2 q_3)_\vma|0\ra\la
V_1|(\bar{q}_1b)_\vma|\ov B \ra \non \\ &=& - if_{V_2}m_2\Bigg[
(\vp^*_1\cdot\vp^*_2) (m_{B}+m_{1})A_1^{ BV_1}(m_{2}^2)  \non \\
&-& (\vp^*_1\cdot p_{_{B}})(\vp^*_2 \cdot p_{_{B}}){2A_2^{
BV_1}(m_{2}^2)\over (m_{B}+m_{1}) } +
i\epsilon_{\mu\nu\alpha\beta}\vp^{*\mu}_2\vp^{*\nu}_1p^\alpha_{_{B}}
p^\beta_1\,{2V^{ BV_1}(m_{2}^2)\over (m_{B}+m_{1}) }\Bigg].
\en

Take the decay $B\to\phi K^*$ as an example. In the naive
factorization approach for hadronic weak decays, the decay
amplitude of $B^-\to\phi K^{*-}$ reads (in units of
$G_F/\sqrt{2}$) \cite{CCTY,AKL}
\begin{eqnarray} \label{famp}
A(B_u^- \to  K^{*-} \phi)= && V_{ub} V^{*}_{us} a_1 X^{(B^-_u
,\phi K^{*-})} - V_{tb} V^{*}_{ts}\Bigg\{ \Big[ a_3+a_4+a_5 \non\\
&& -{1\over2}(a_7+a_9+a_{10})\Big] X_s^{( B^-  K^{*-},\phi)}
+(a_4+a_{10}) X^{( B^-,  K^{*-}\phi)}\nonumber\\ &&-2(a_6+a_8)
\langle  K^{*-}\phi|\bar s(1+\gamma_5)u|0\rangle \langle 0|\bar
u(1-\gamma_5)b| B^-\rangle \Bigg\},
\end{eqnarray}
where $a_{2i}=c_{2i}+{1\over N_c}c_{2i-1},~
a_{2i-1}=c_{2i-1}+{1\over N_c}c_{2i}$. Neglecting the annihilation
contributions from the last two terms in Eq. (\ref{famp}), we
obtain
\be
H_{00} &=& {\tilde a(\phi K^*)f_\phi\over
2m_{K^*}}\left[(m_B^2-m_{K^*}^2-m_\phi^2)(m_B+m_{K^*})A^{BK^*}_1(m_\phi^2)
-{4m_B^2p_c^2\over m_B+m_{K^*} }\,A^{BK^*}_2(m_\phi^2)\right],
\non \\ H_{\pm\pm} &=& \tilde a(\phi K^*)m_\phi f_\phi\left[
(m_B+m_{K^*})A_1^{BK^*}(m_\phi^2)\mp {2m_Bp_c\over m_B+m_{K^*}}
V^{BK^*}(m_\phi^2)\right],
\en
where $\tilde a(\phi K^*)=a_3 + a_4 + a_5 -{1\over 2}( a_7+ a_9 +
a_{10})$. In the heavy quark limit, it is clear that
\be
H_{00} &=& {\tilde a(\phi K^*)f_\phi m_B^3\over
2m_{K^*}}[A^{BK^*}_1(0)- A^{BK^*}_2(0)],  \non \\ H_{\pm\pm} &=&
\tilde a(\phi K^*)f_\phi m_\phi m_B[A_1^{BK^*}(0)\mp V^{BK^*}(0)].
\label{phiK}
\en

In the so-called generalized factorization, nonfactorizable
effects are parametrized in terms of $\nc$, the effective number
of colors. This amounts to assuming that nonfactorizable
corrections weight in the same way to all partial-wave or helicity
amplitudes. For example, the coefficient $\tilde a(\phi K^*)$
appearing in Eq. (\ref{phiK}) is postulated to be the same for
$S$, $P$ and $D$ (or $H_{00}$ and $H_{\pm\pm}$) amplitudes after
including nonfactorizable contributions. Clearly there is no any
known physical argument for justifying this assumption.

Fortunately, the QCD-improved factorization approach advocated
recently in \cite{BBNS} allows us to compute the nonfactorizable
corrections in the heavy quark limit since only hard interactions
between the $(BV_1)$ system and $V_2$ survive in the
$m_b\to\infty$ limit. Naive factorization is recovered in the
heavy quark limit and to the zeroth order of QCD corrections. In
this approach, the light-cone distribution amplitudes (LCDAs) play
an essential role. The LCDAs of the light vector meson of interest
are given by \cite{Ballv,BBNS}
\be
\la V(p,\vp)|\bar
q_\alpha(y)q'_\beta(x)|0\ra  &=& {f_Vm_V\over 4}
\int^1_0du\,e^{i(up\cdot y+\bar u p\cdot
x)}\Bigg[\vp\!\!\!/^*_\|\Phi^V_\|(u)+\vp\!\!\!/^*_\bot
g_\bot^{(v)}(u) \non
\\&& +{1\over 4}\Big(1-{f_V^T\over f_V}\,{m_{q_1}+m_{q_2}\over
m_V}\Big) \epsilon
_{\mu\nu\rho\sigma}\gamma^\mu\gamma_5\vp^{*\nu}p^\rho z^\sigma
g_\bot^{(a)}(u)\Bigg]_{\alpha\beta} \non
\\ && +{f_V^T\over 4} \,\int^1_0du\,e^{i(up\cdot y+\bar u p\cdot
x)}(\vp\!\!\!/^*_\bot p\!\!\!/)_{\alpha\beta}\Phi^V_\bot(u),
\label{Jwf}
 \en
where $z=y-x$ with $z^2=0$, $\vp^\mu_\|$ ($\vp^\mu_\bot$) is the
polarization vector of a longitudinally (transversely) polarized
vector meson, $u$ is the light-cone momentum fraction of the quark
$q$ in the vector meson, $\bar u=1-u$, $f_V$ and $f^T_V$ are
vector and tensor decay constants, respectively, but the latter is
scale dependent. To a good approximation one has
$\vp^\mu_\|=p^\mu_V/m_V$ for a light vector meson. It follows that
$p\cdot \vp_\bot=0$. It is easily seen from Eq. (\ref{amp}) that
$\Phi_\|$ contributes to $S$ and $D$ amplitudes, while $\Phi_\bot$
to $P$ and $S$ waves. In Eq. (\ref{Jwf}), $\Phi_\|(u)$ and
$\Phi_\bot(u)$ are twist-2 DAs, while $g_\bot^{(v)}$ and
$g_\bot^{(a)}$ are twist-3 ones. Note that twist-3 longitudinally
polarized distribution amplitudes $h_\|^{(s)}$ and $h_\|^{(t)}$
\cite{Ballv} are not shown in Eq. (\ref{Jwf}). The reason for
keeping the twist-3 transversely polarized distribution amplitudes
$g_\bot^{(v,a)}$ rather than the longitudinal ones $h_\|^{(s,t)}$
will become clear shortly.

\vskip 0.7 cm {\bf 3.}~~We will now study charmless $B\to VV$
decays within the framework of QCD-improved factorization. The
power corrections such as the annihilation diagrams can be
neglected in the heavy quark limit. It turns out that the twist-2
DA $\Phi_\bot(u)$ contributions to the vertex corrections and hard
spectator interactions vanish in the chiral limit.\footnote{For
the color-suppressed mode $B\to J/\psi K^*$, the twist-2
transverse polarized distribution amplitude is on the same footing
as the longitudinal one since $J/\psi$ is not massless in heavy
quark limit.} Hence, we will work to the leading-twist
approximation for longitudinally polarized states and to the
twist-3 level for the case of transverse polarization. As
discussed before, the effective parameters $a_i$ entering into the
helicity amplitudes $H_{00}$ and $H_{\pm\pm}$ are not the same;
they are given by
\be \label{ai}
 a_1^h &=& c_1+{c_2\over N_c}+{\alpha_s\over 4\pi}\,{C_F\over N_c}c_2\,F^h,
\non \\
 a_2^h &=& c_2+{c_1\over N_c}+{\alpha_s\over 4\pi}\,{C_F\over
N_c}c_1\,F^h, \non \\
 a_3^h &=& c_3+{c_4\over N_c}+{\alpha_s\over
4\pi}\,{C_F\over N_c} c_4\,F^h, \non \\
 a_4^h &=& c_4+{c_3\over N_c}+{\alpha_s\over 4\pi}\,{C_F\over
N_c}\Bigg\{c_3\big[F^h+G^h(s_q)+G^h(s_b)\big]-c_1\left({\lambda_u\over
\lambda_t}G^h(s_u)+{\lambda_c\over\lambda_t}G^h(s_c)\right)   \non
\\ && +(c_4+c_6) \sum_{i=u}^b G^h(s_i)  +{3\over 2} (c_8+c_{10})\sum
_{i=u}^b e_i G^h(s_i)+{3\over 2}c_9\big[e_qG^h(s_q)-{1\over
3}G^h(s_b)\big]+c_g G^h_g\Bigg\}, \non \\
 a_5^h &=& c_5+{c_6\over
N_c}+{\alpha_s\over 4\pi}\,{C_F\over N_c} c_6(-F^h-12), \\
 a_6^h &=& c_6+{c_5\over N_c}, \non \\
 a_7^h &=& c_7+{c_8\over N_c}+{\alpha_s\over 4\pi}\,{C_F\over N_c}
c_8(-F^h-12)-{\alpha\over 9\pi}\,N_cC^h_e,  \non \\
 a_8^h &=& c_8+{c_7\over N_c},  \non \\
 a_9^h &=& c_9+{c_{10}\over N_c}+{\alpha_s\over 4\pi}\,{C_F\over N_c}
 c_{10}\,F^h-{\alpha\over 9\pi}\,N_cC^h_e,  \non \\
 a_{10}^h &=& c_{10}+{c_9\over
N_c}+{\alpha_s\over 4\pi}\,{C_F\over N_c} c_9\,F^h-{\alpha\over
9\pi}\,C^h_e, \non
\en
 where $C_F=(N_c^2-1)/(2N_c)$, $s_i=m_i^2/m_b^2$,
$\lambda_{q}= V_{qb}V^*_{qq'}$, $q'=d,s$ and the superscript $h$
denotes the polarization of the vector mesons: $h=0$ for helicity
00 states, and $h=\pm$ for helicity $\pm\pm$ states.

There are QCD penguin-type diagrams induced by the 4-quark
operators $O_i$ for $i=1,3,4,6,8,9,10$. The corrections are
described by the penguin-loop function $G^h(s)$ given by
\be
G^0(s) &=& {2\over 3}-{4\over 3}\ln{\mu\over m_b}+4\int^1_0
dx\,\Phi^{V}_\|(x)\int^1_0 du\,u(1-u)\ln[s-x u(1-u)], \non \\
G^\pm(s) &=& {2\over 3}-{4\over 3}\ln{\mu\over m_b}+4\int^1_0
dx\,g^{(v)}_\bot(x)\int^1_0 du\,u(1-u)\ln[s-x u(1-u)]  \\ &\mp&
{1\over 2}\int^1_0 dx\,{g_\bot^{(a)}(x)\over x}\int^1_0
du\,u(1-u)\Bigg\{-2 \ln{\mu\over m_b}+\ln[s-x u(1-u)]+{x
u(1-u)\over s-x u(1-u)}\Bigg\}. \non \label{G} \en In Eq.
(\ref{ai}) we have also included the leading electroweak
penguin-type diagrams induced by the operators $O_1$ and $O_2$
\cite{Ali}: \be C_e^h &=& \left({\lambda_u\over
\lambda_t}G^h(s_u)+{\lambda_c\over
\lambda_t}G^h(s_c)\right)\left(c_2+{c_1\over N_c}\right). \en The
dipole operator $O_g$ will give a tree-level contribution
proportional to \be G^0_g = -2\int^1_0dx\,{\Phi^{V}_\|(x)\over x},
\qquad G^\pm_g = -2\int^1_0dx\,{g^{(v)}_\bot(x)\over x}+{1\over
2}(1\mp {1\over 2}) \int^1_0dx\,{g^{(a)}_\bot(x)\over x}. \en

In Eq. (\ref{ai}), the vertex correction in the naive dimensional
regularization (NDR) scheme for $\gamma_5$ is given by \be
F^h=-12\ln{\mu\over m_b}-18+f_I^h+f_{II}^h, \label{F} \en where
the hard scattering function $f_I$ arises from vertex corrections
and $f_{II}$ from the hard spectator interactions with a hard
gluon exchange between the emitted vector meson and the spectator
quark of the $B$ meson. Note that the twist-2 transversely
polarized distribution amplitude $\Phi_\bot$ does not contribute
to $F$: it does not give rise to the scale and scheme dependent
terms $-12\ln(\mu/m_b)-18$ and the hard scattering kernels
$f_I^\pm$ and $f_{II}^\pm$ are proportional to the light quark
mass and hence can be neglected. Therefore, the parameters
$a_i^\pm$ at the twist-2 level are not renormalization scale and
scheme independent. Consequently, it is necessary to take into
account the twist-3 effects for transversely polarized vector
meson states. This is why we keep twist-3 DAs $g_\bot^{(a,v)}$ in
Eq. (\ref{Jwf}). As for the helicity zero case, it is dominated by
the leading-twist one and hence the twist-3 DAs $h_\|^{(s,t)}$,
which are power suppressed by order of $m_V/m_B$, are not
considered there. It should be stressed that $a_{6,8}^h$ do depend
on the choice of the renormalization scale and scheme. Their scale
and scheme dependence is compensated by the corresponding
$(S-P)(S+P)$ hadronic matrix elements.

An explicit calculation for $f_I^h$ yields
\be
f_I^0=\int^1_0dx\,\Phi_\|^{V}(x)\left(3{1-2x\over 1-x}\ln
x-3i\pi\right), \non \\
f_I^\pm=\int^1_0dx\,g_\bot^{(v)}(x)\left(3{1-2x\over 1-x}\ln
x-3i\pi\right), \label{fI}
\en
where $f_I^0$ has the same expression as the hard scattering
kernel $f_I$ in $B\to\pi\pi$ \cite{BBNS}. The hard kernel
$f_{II}^h$  for hard spectator interactions have the expressions
($V_1$: recoiled meson, $V_2$: emitted meson):
\be
f_{II}^0 &=& {4\pi^2\over N_c}\,{2f_Bf_{V_1}m_1\over
h_{0}}\int^1_0 d\drho\, {\Phi^B_1(\drho)\over \drho}\int^1_0
d\deta \,{\Phi^{V_1}_\|(\deta)\over \deta}\int^1_0 d\xi\,
{\Phi^{V_2}_\|(\xi)\over \xi}, \non \\
 f_{II}^\pm &=& -
{4\pi^2\over N_c}\,{f_Bf^T_{V_2}\over m_Bh_{\pm}}2(1\mp 1)\int^1_0
d\drho\, {\Phi^B_1(\drho)\over \drho}\int^1_0 d\deta\,
{\Phi^{V_1}_\bot(\deta)\over
\deta^2}\int^1_0d\xi\,g_\bot^{V_2(v)}(\xi) \non \\  && +
{4\pi^2\over N_c}\,{2f_Bf_{V_1}m_1\over m_B^2h_{\pm}}\int^1_0
d\drho\, {\Phi^B_1(\drho)\over\drho}\int^1_0 d\deta\,d\xi\Bigg\{
g^{V_1(v)}_\bot(\deta)g_\bot^{V_2(v)}(\xi)\,{\xi+\deta\over
\xi\,\deta^2} \non \\ & & \pm{1\over 4}
g^{V_1(v)}_\bot(\deta)g_\bot^{V_2(a)}(\xi)\,{\xi+\deta\over
\xi^2\,\deta^2}\mp{1\over 4}
g^{V_1(a)}_\bot(\deta)g_\bot^{V_2(v)}(\xi)\,{2\xi+\deta\over
\xi\deta^3}\Bigg\},
 \label{fII2}
\en
where
\be
h_{0} &=& (m_B^2-m_{1}^2-m_{2}^2)(m_B+m_{1})A^{BV_1}_1(m_2^2)
-{4m_B^2p_c^2\over m_B+m_1 }\,A^{BV_1}_2(m_2^2), \non \\ h_{\pm}
&=& (m_B+m_{1})A^{BV_1}_1(m_2^2)\mp {2m_Bp_c\over m_B+m_1
}\,V^{BV_1}(m_2^2),
\en
and  we have neglected the light quark masses and applied the
approximation $\bar\rho\approx 0$, and the $B$ meson wave function
\cite{BBNS}:
\be
\la 0|\bar q_\alpha(x)b_\beta(0)|\bar
B(p)\ra\!\!\mid_{x_+=x_\bot=0}=-{if_B\over 4}[(p\!\!\!/
+m_B)\gamma_5]_{\beta\gamma}\int^1_0d\drho\, e^{-i\drho
p_+x_-}[\Phi^B_1(\drho)+n\!\!\!/_-\Phi^B_2(\drho)]_{\gamma\alpha},
 \label{Bwf}
\en with  $n_-=(1,0,0,-1)$. Note that the presence of logarithmic
and linear infrared divergences in $f_{II}^\pm$ implies that the
spectator interaction is dominated by soft gluon exchanges in the
final states. Hence, factorization breaks down at the twist-3
order for transversely polarized vector meson states. We will
introduce a cutoff of order $\Lqcd/m_b$ to regulate the linear and
logarithmic divergences. The choice of the cutoff is not important
here since the transversely polarized amplitudes are suppressed
anyway.

Two remarks are in order. (i) Since $\la V|\bar q_1q_2|0\ra=0$,
$B\to VV$ decays do not receive factorizable contributions from
$a_6$ and $a_8$ penguin terms except for spacelike penguin
diagrams \cite{CCTY}. (ii) The first two terms
$-12\ln(\mu/m_b)-18$ in Eq. (\ref{F}) for helcity $\pm\pm$ states
arise from the twist-3 DA $g_\bot^{(v)}(u)$ and will render the
parameters $a_i^\pm$ (except for $a_6^\pm$ and $a_8^\pm$) scale
and scheme independent.

\vskip 0.7cm
 {\bf 4.}~~To proceed we compute the branching ratios using LCSR
and BSW models for heavy-light form factors (see Table I). The
factorized amplitudes of $B\to VV$ modes are given in
\cite{CCTY,AKL}. Note that the original BSW model assumes a
monopole behavior for all the form factors. This is not consistent
with heavy quark symmetry for heavy-to-heavy transition.
Therefore, we will employ the BSW model for the heavy-to-light
form factors at zero momentum transfer but take a different ansatz
for their $q^2$ dependence, namely a dipole dependence for
$A_0,A_2$ and $V$. In the light-cone sum rule analysis, the
form-factor $q^2$ dependence is given in \cite{Ball}.

To proceed we use the next-to-leading Wilson coefficients in the
NDR scheme \cite{Buras96}
\be
c_1=1.082, \quad c_2=-0.185,\quad c_3=0.014, \quad c_4=-0.035,
\quad c_5=0.009, \quad c_6=-0.041,  \non \\ c_7/\alpha=-0.002,
\quad c_8/\alpha=0.054, \quad c_9/\alpha=-1.292, \quad
c_{10}/\alpha=0.263,\quad c_g=-0.143,
\en
with $\alpha$ being an electromagnetic fine-structure coupling
constant. For the LCDAs, we use the asymptotic form for the vector
meson \cite{Ballv}
\be
&& \Phi^V_\|(x)=\Phi^V_\bot(x)=g_\bot^{(a)}(x)=6x(1-x),  \non \\
&& g_\bot^{(v)}(x)={3\over 4}\left[ 1+(2x-1)^2\right],
\en
and the $B$ meson wave function
\be
\Phi^B_1(\drho)=N_B\drho^2(1-\drho)^2{\rm exp}\left[-{1\over
2}\left({\drho m_B\over \omega_B}\right)^2\right], \label{Bda}
\en
with $\omega_B=0.25$ GeV and $N_B$ being a normalization constant.
For the decay constants, we use
\be
 f_\rho=216\,{\rm MeV},\qquad
f_{K^*}=221\,{\rm MeV}, \qquad f_\omega=195\,{\rm MeV},\qquad
f_\phi=237\,{\rm MeV},
\en
and we will assume $f^T_V=f_V$ for the tensor decay constant.

\vskip 0.4cm
\begin{table}[ht]
\caption{Form factors at zero momentum transfer for $B\to P$ and
$B\to V$ transitions evaluated in the light-cone sum rule (LCSR)
analysis [3]. The values given in the square brackets are obtained
in the BSW model [2]. We have assumed SU(3) symmetry for the
$B\to\omega$ form factors in the LCSR approach.}
\begin{center}
\begin{tabular}{ l c c c c }
Decay  & $V$ & $A_1$ & $A_2$ & $A_3=A_0$ \\ \hline
 $B\to\rho^\pm$ & 0.338~[0.329] & 0.261~[0.283] & 0.223~[0.283] &
 0.372~[0.281] \\
 $B\to\omega$ & 0.239~[0.232] & 0.185~[0.199] & 0.158~[0.199] &
 0.263~[0.198] \\
 $B\to K^*$  & 0.458~[0.369] & 0.337~[0.328] &
 0.283~[0.331] & 0.470~[0.321] \\
\end{tabular}
\end{center}
\end{table}

\vskip 0.4cm
\begin{table}[ht]
\caption{Branching ratios (in units of $10^{-6}$) averaged over
CP-conjugate modes for charmless $B\to VV$ decays. Two different
form-factor models, the LCSR and the BSW models, are adopted and
the unitarity angle $\gamma=60^\circ$ is employed. Experimental
limits and results are taken from [14-16,4-6 and the limits
indicated by * are quoted from [14] only for the helicity zero
states.}
\begin{center}
\begin{tabular}{l l c l }
Decay & LCSR & BSW & Expt. \\ \hline
 $\ov B^0 \to \rho^-  \rho^+$
  & 35.0 & 21.2 &  $ <2200$  \\
 $\ov B^0 \to \rho^0 \rho^0$
  & 0.26 & 0.20 & $<5.9^*$  \\
 $ \ov B^0 \to \omega\,\omega$
  & 0.30 & 0.22 & $ <19$\\
 $B^-\to  \rho^- \rho^0$
  & 21.8 & 12.8 & $<120$\\
 $ B^- \to \rho^-\omega$
  & 21.0 & 13.8 & $<47$\\
 $\ov B^0 \to K^{*-} \rho^+$
  & 4.84 & 3.12 & $ -$\\
 $\ov B^0 \to\ov K^{*0}\rho^0$
  & 0.99 & 0.71 & $<19^* $ \\
 $\ov B^0 \to\ov K^{*0}K^{*0}$
  & 0.32 & 0.16 & $<10^*$\\
 $ B^- \to K^{*- }\rho^0$
  & 5.59 & 2.94 & $ <54^*$\\
 $ B^- \to \ov K^{*0} \rho^-$
  & 6.70 & 4.01 &   $-$\\
 $ B^- \to K^{*- } K^{*0}$
  & 0.34 & 0.15 &  $<50^*$\\
 $ \ov B^0 \to \rho^{0} \phi$
  &0.003 & 0.002 & $ <13$\\
 $ \ov B^0 \to\omega\, \phi$
  & 0.003 & 0.002& $ <21$\\
 $ B^- \to\rho^{-} \phi$
  & 0.007 & 0.004 &  $ <16$\\
 $ \ov B^0 \to \rho^0 \omega$
  & 0.12 & 0.07 & $ <11$\\
$\ov B^0 \to \ov K^{*0}\omega$
  & 3.66 & 2.15 &  $<19$\\
   $ B^- \to K^{*-} \omega $
  & 3.12 & 1.88 &   $ <52$\\
  $ B^- \to K^{*-} \phi$
  & 9.30 & 4.32 & $9.7^{+4.2}_{-3.4}\pm1.7$ \cite{BaBar} \\
  & & & $10.6^{+6.4+1.8}_{-4.9-1.6}$ \cite{Briere}\\
  & & & $<36$ \cite{BELLE} \\
   $ \ov B^0 \to\ov K^{*0} \phi$
  & 8.71 & 4.62 & $8.6^{+2.8}_{-2.4}\pm1.1$ \cite{BaBar}\\
  & & & $11.5^{+4.5+1.8}_{-3.7-1.7}$ \cite{Briere}\\
  & & & $15^{+8}_{-6}\pm 3$ \cite{BELLE} \\
\end{tabular}
\end{center}
\end{table}

To illustrate the non-universality of nonfactorizable effects for
helicity amplitudes, we give a few numerical results for the
parameters $a_i^h$:
 \be
 && a_1^0=1.04+0.01i, \qquad a_1^+=1.02+0.01i, \qquad
 a_1^-=1.11+0.04i,  \non \\
  && a_2^0=0.09-0.08i, \qquad a_2^+=0.17-0.08i, \qquad
 a_2^-=-0.38-0.25i,  \non \\
  && a_4^0=-0.033-0.004i, \qquad a_4^+=-0.026-0.004i, \qquad
 a_4^-=-0.040-0.007i,
 \en
in the LCSR model for form factors. Therefore, nonfactorizable
corrections to helicity amplitudes are not universal. From Table
II we see that the branching ratios predicted by LCSR is larger
than that by the BSW model by a factor of $1.5\sim 2$. Evidently,
the experimental results for $B\to\phi K^*$ favor the LCSR form
factors for $B-V$ transition. It should be stressed that thus far
we have not taken into account power corrections such as
annihilation diagrams and higher-twist wave functions for the
longitudinally polarized vector meson. In particular, weak
annihilations induced by the $(S-P)(S+P)$ penguin operators are no
longer subject to helicity suppression and hence can be sizable
(see the last term in Eq. (\ref{famp}) and \cite{CYphiK}).
However, contrary to the $PP$ and $PV$ modes, the annihilation
amplitude in the $VV$ case does not gain a chiral enhancement of
order $m_B^2/(m_qm_b)$. Therefore, it is truly power suppressed in
the heavy quark limit.

It is also clear from Table II  that the tree-dominated modes
$\rho^+\rho^-,~\rho^-\rho^0,~\rho^-\omega$ have larger branching
ratios of order $(2\sim 3)\times 10^{-5}$ than the
penguin-dominated ones. This is ascribed to the fact that the
$(S-P)(S+P)$ penguin operators do not contribute to factorizable
$W$-emission amplitudes. By contrast, the $\rho^0\rho^0$ and
$\omega\omega$ modes have rather small branching ratios because
the parameter $a_2$ is small in QCD factorization.  We have also
computed $|H_{00}|^2$ and $|H_{\pm\pm}|^2$ for each channel and
found that $|H_{--}/H_{00}|^2=(5\sim 20)\%$ and
$|H_{++}/H_{00}|^2=(10^{-5}\sim 10^{-3})$.

\vskip 0.7cm
 {\bf 5.}~~We have analyzed $B\to VV$ decays within the framework
of QCD factorization. Contrary to phenomenological generalized
factorization, nonfactorizable corrections to each partial-wave or
helicity amplitude are not the same; the effective parameters
$a_i$ vary for different helicity amplitudes. The leading-twist
nonfactorizable corrections to the transversely polarized
amplitudes vanish in the chiral limit and hence it is necessary to
take into account twist-3 distribution amplitudes of the vector
meson in order to have renormalization scale and scheme
independent predictions. Branching ratios of $B\to VV$ decays are
calculated in two different models for form factors, and the
predicted decay rates are different by a factor of $1.5\sim 2$. In
the heavy quark limit, both vector mesons in the charmless $B\to
VV$ decay should have zero helicity and the corresponding
amplitude is proportional to the form factor difference
$(A_1-A_2)$. The recent observation of $B\to\phi K^*$ indicates
that the form factors $A_1(q^2)$ and $A_2(q^2)$ cannot be very
similar at low $q^2$ as implied by the BSW model. Owing to the
absence of $(S-P)(S+P)$ penguin operator contributions to
$W$-emission amplitudes, tree-dominated $B\to VV$ decays tend to
have larger branching ratios than the penguin-dominated ones.

\vskip 2 cm  \acknowledgments  One of us (H.Y.C.) wishes to thank
Physics Department, Brookhaven National Laboratory for its
hospitality. This work was supported in part by the National
Science Council of R.O.C. under Grant Nos. NSC89-2112-M-001-082
and NSC89-2112-M-033-014.


\end{document}